\documentclass[conference,10pt]{IEEEtran}
\usepackage{epsfig,rotating,setspace,latexsym,amsmath,epsf,amssymb,amsfonts,bm,theorem,subfigure,epstopdf}
\usepackage{cite,authblk}
\usepackage{bbm}
\usepackage{color, xcolor}
\usepackage{mathtools}
\usepackage{algorithm}
\usepackage{textcomp}
\usepackage[noend]{algpseudocode}
\usepackage{verbatim}
\usepackage{graphicx, graphics}
\usepackage{afterpage}
\usepackage{placeins}
\usepackage{bm}
\usepackage{authblk}
\usepackage[symbol]{footmisc}

\DeclareMathOperator*{\argmax}{arg\,max}

\algrenewcommand\algorithmicforall{\textbf{foreach}}
\algrenewcommand\algorithmicindent{.8em}

\IEEEoverridecommandlockouts
\allowdisplaybreaks

\begin{document}

\title{Deep Learning Based Uplink Multi-User SIMO Beamforming Design}

\author{
Cemil Vahapoglu$^{\dag}$, Timothy J. O’Shea$^{*}$, Tamoghna Roy$^{*}$, Sennur Ulukus$^{\dag}$ \\
\normalsize $^{\dag}$University of Maryland, College Park, MD, $^{*}$DeepSig Inc., Arlington, VA \\
\normalsize \emph{cemilnv@umd.edu, tim@deepsig.io, tamoghna.roy@deepsig.io, ulukus@umd.edu}
}

\maketitle

\begin{abstract}
The advancement of fifth generation (5G) wireless communication networks has created a greater demand for wireless resource management solutions that offer high data rates, extensive coverage, minimal latency and energy-efficient performance. Nonetheless, traditional approaches have shortcomings when it comes to computational complexity and their ability to adapt to dynamic conditions, creating a gap between theoretical analysis and the practical execution of algorithmic solutions for managing wireless resources. Deep learning-based techniques offer promising solutions for bridging this gap with their substantial representation capabilities. We propose a novel unsupervised deep learning framework, which is called NNBF, for the design of uplink receive multi-user single input multiple output (MU-SIMO) beamforming. The primary objective is to enhance the throughput by focusing on maximizing the sum-rate while also offering computationally efficient solution, in contrast to established conventional methods. We conduct experiments for several antenna configurations. Our experimental results demonstrate that NNBF exhibits superior performance compared to our baseline methods, namely, zero-forcing beamforming (ZFBF) and minimum mean square error (MMSE) equalizer. Additionally, NNBF is scalable to the number of single-antenna user equipments (UEs) while baseline methods have significant computational burden due to matrix pseudo-inverse operation.
\end{abstract}

\section{Introduction}

Wireless physical layer research has generally focused on waveform design, channel characterization, and the developments of signaling and detection techniques, including tasks such as interference management, design of transmitter and receiver chains, and implementing error correction algorithms to provide reliable data transfer \cite{OsheaDL}. With the evolution of fifth generation (5G) wireless communication networks, a greater demand for high data rates, large coverage, low latency and power efficiency has arisen. These problems of interest can be considered within the scope of wireless resource management problems that cover a range of areas including power control, spectrum management, backhaul management, computation resource management, cache management, transmit/receive beamforming design and so forth \cite{AppML}.

The design and implementation of  conventional wireless communication systems require strong probabilistic modeling and signal processing techniques \cite{erpek2020deep}. However, existing techniques in wireless communications have formidable limitations in terms of computational complexity since they necessitate rigorous computations. This causes an important gap between theoretical analysis and real-time processing of algorithms to aforementioned problems of wireless resource management. Aside from the significant computational complexity, many of the existing designs are unsuitable for dynamic network scenarios due to their presumptions of static network conditions and inability to utilize the dynamic data effectively \cite{AppML}. 

On the contrary, machine learning (ML) offers robust automated systems that can learn adaptively from dynamic spectrum data instead of relying on solely policy based solutions for specific scenarios \cite{ClancyCognitive}. Recent developments of fast and powerful graphical processing units (GPUs) and the significant growth of accessible data allowed especially deep learning-based methods to achieve considerable representation capabilities \cite{AppML}. Moreover, it has been proven that the deep neural network (DNN) provides universal function approximations for traditional high complexity algorithms \cite{Sun2018}.  Authors in \cite{Sun2018} suggest that numerical optimization problems addressing wireless resource management problems such as beamforming design and power control can be evaluated as a non-linear mapping function to be learned by a DNN.

In addition to the function approximation by DNN, there are also studies that focus on approximating iterative optimization algorithms by unfolding iterations using DNN. In other words, deep unfolding algorithms imitate the iterations through layers that correspond to a neural network \cite{Hershey2014DeepUnfolding}. Namely, \cite{Gregor2010LearningFA} proposes multilayer network to approximate iterative soft-thresholding algorithm for sparse optimization. Similarly, \cite{Samuel2017DeepMIMO} proposes a neural network architecture by unfolding a projected gradient descent method for the design of robust MIMO detector against ill-conditioned channel with lower computation complexity. However, deep unfolding approach may not be feasible for complex algorithms. Wireless resource management algorithms usually require  matrix inversion and singular value decomposition (SVD), both of which involve computationally heavy iterations for approximation \cite{Sun2018}.

The focus of this paper is the beamforming design, which is a significant challenge within 5G wireless communication networks. In the literature, several deep learning based beamforming design methods have been proposed for different MIMO configurations. \cite{WenchaoMISODownlinkBF} proposes a deep learning model based on the convolutional neural network (CNN) and expert knowledge to utilize the structure of optimal solutions for signal-to-interference-plus-noise ratio (SINR) balancing problem, power minimization problem, and sum-rate maximization problem. They use the minimal transmit power satisfying the constraints in power minimization problem to design downlink beamforming vectors to achieve maximal sum-rate as suggested by \cite{BjornsonOptimalResource2013}. However, the computation of downlink beamforming vectors from minimal transmit power also involves matrix inversion, which can create computational burden for a real-time processing system when the system has very large number of antennas as in massive MIMO. In \cite{Huang2018FastBF}, authors propose a fully connected network for downlink beamforming design to maximize the weighted sum-rate under a total power constraint. \cite{Zhang_2023} proposes a joint learning framework for channel prediction, power optimization, and transmit beamforming prediction in the downlink channel. However, it also exploits the parameterized structure of beamforming solution for sum-rate maximization problem given the power values as result of power minimization problem as in \cite{WenchaoMISODownlinkBF}. Additionally, \cite{DeepTx2022} proposes a CNN to design downlink beamforming by uplink channel estimate in a supervised manner. 

Apart from utilizing deep learning for downlink beamforming design, there are also other works that propose deep learning-based methods for fully learned receivers on the uplink channel to recover uncoded bits through supervised learning, with a primary focus on leveraging deep learning for channel estimation \cite{Korpi2020DeepRxMC, PowerDL2018, DeepWaveform2021}.

In our work, we introduce a novel deep learning-based approach for the design of uplink receive multi-user single input multiple output (MU-SIMO) beamforming. The proposed framework is denoted as NNBF, which is abbreviated for neural network beamforming throughout the paper. NNBF performs unsupervised training. To the best of our knowledge, it is the first work that utilizes unsupervised deep learning training for the purpose of uplink receive beamforming design, targeting the sum-rate maximization problem. We conduct the performance analysis of our proposed framework by comparing with zero-forcing  beamforming (ZFBF) technique and minimum mean square error (MMSE) equalizer, which are considered as our baselines. 

The proposed framework surpasses the performance of ZFBF across varying number of receive antennas while it has competitive performance with MMSE baseline. Moreover, experiments are conducted to investigate NNBF performance for different SNR regimes as the network size expands while maintaining a constant ratio of single-antenna user equipments (UEs) to the number of receive antennas. Results exhibit that NNBF performs comparable or better than MMSE baseline while it always surpasses ZFBF performance for all SNR regimes. Furthermore, the experiments demonstrate the ability of the proposed framework to scale as the number of single-antenna UEs increases when baseline methods have significant increase in computation time due to the matrix pseudo-inversion operation.

\section{System Model and Problem Formulation}
\subsection{Uplink Multi-User SIMO (MU-SIMO) Setup}
\begin{figure}[t]
 \centerline{\includegraphics[width= 1\linewidth]{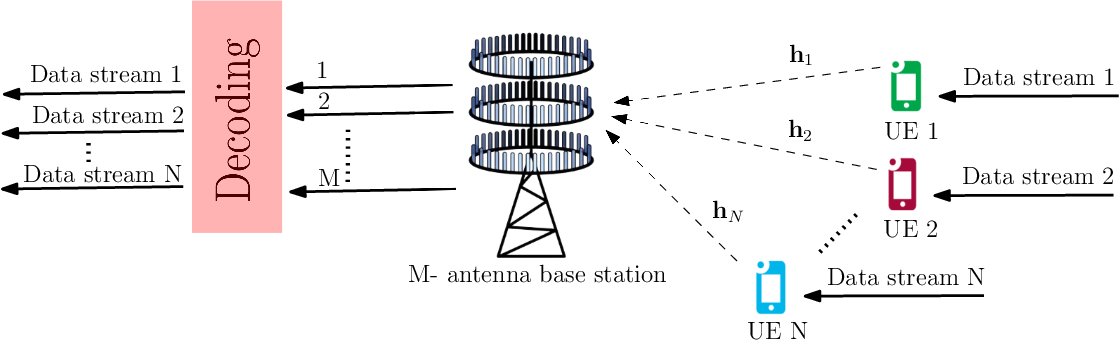}}
  \caption{Uplink massive MIMO link where UEs transmit data streams on the same time/frequency resources.}
  \label{System Model}
  \vspace*{-0.1cm}
\end{figure}

We consider an uplink transmission scenario where $N$ single-antenna UEs transmit data streams to a base station (BS) equipped with $M$ receive antennas as shown in Fig~\ref{System Model}. 

The uplink channel matrix is denoted as $\mathbf{H} = 
[\mathbf{h}_1  ~ \mathbf{h}_2 ~ \cdots ~ \mathbf{h}_N] \in \mathbb{C}^{M \times N}$, where $\mathbf{h}_k$ corresponds to the channel vector between UE $k$ and the BS. Then, the received signal at the BS can be written as
\begin{align} \label{received_signal}
    \mathbf{y} = \sum_{i=1}^N \mathbf{h}_i x_i +\mathbf{n} 
\end{align}
$\mathbf{x} = [ x_1^H ~ x_2^H ~ \cdots ~ x_N^H]^H \in \mathbb{C}^N$ is the transmitted signal whose entries are the data symbols transmitted by UEs with $\mathbb{E}[x_k^Hx_k] =1, \, \forall k=1,\ldots, N.$ Additionally, $\mathbf{n} = [ n_1^H ~ n_2^H ~ \cdots ~ n_M^H]^H \in \mathbb{C}^M$ denotes the additive white Gaussian noise with i.i.d.~entries $n_l \sim \mathcal{CN}(0, \sigma^2), \, \forall l=1,\ldots, M$.

It is assumed that channel state information (CSI) is available on the BS. Then, the received signal in (\ref{received_signal}) is processed by beamforming weights $\mathbf{W} = [\mathbf{w}_1  ~ \mathbf{w}_2 ~ \cdots ~ \mathbf{w}_N
]^T \in \mathbb{C}^{N \times M}$ to recover data streams, where power normalization of receive antennas is satisfied as $\mathrm{tr}(\mathbf{W}^H\mathbf{W}) = M$. Specifically, $\mathbf{w}_k \in \mathbb{C}^{M}$ represents the linear beamforming filter to estimate the transmitted data symbols of UE $k$,
\begin{align}\label{received_uek}
    \mathbf{w}_k^T \mathbf{y} & = \sum_{i=1}^N \mathbf{w}_k^T\mathbf{h}_i x_i +\mathbf{w}_k^T\mathbf{n}
\end{align}

\subsection{Beamforming Design for Sum-rate Maximization Problem}

Our objective is to design beamforming weights to maximize the sum-rate of all UEs. For UE $k$, the received signal with beamforming weight $w_k$ given in (\ref{received_uek}) can be rewritten as
\begin{align} \label{received_uek_v2}
    \tilde{x}_k &= \mathbf{w}_k^T \mathbf{y} \nonumber \\ 
     &= \underbrace{\mathbf{w}_k^T\mathbf{h}_k x_k}_{desired \, signal} + \underbrace{ \sum_{i=1, i\neq k}^N \mathbf{w}_k^T\mathbf{h}_i x_i}_{interfering \, signal} +\underbrace{\mathbf{w}_k^T\mathbf{n}}_{noise}
\end{align}
The first term in (\ref{received_uek_v2}) is the desired symbol of UE $k$ while the second term and the third term represent the inter-symbol interference (ISI) of other UEs and the receiver noise, respectively. Then, SINR for UE $k$ can be written as
\begin{align}\label{sinr}
    \gamma_k = \frac{|\mathbf{w}_k^T \mathbf{h}_k|^2}{\sum_{i=1, i\neq k}^N |\mathbf{w}_k^T\mathbf{h}_i|^2 + \mathbb{E}|\mathbf{w}_k^T \mathbf{n}|^2}
\end{align}
Therefore, beamforming design for sum-rate maximization is
\begin{align}\label{sum-rate maximizatiob problem}
    \mathbf{W}^* = \argmax_{\mathbf{W}} & \quad \sum_{i=1}^N \alpha_i \log(1 + \gamma_i) \nonumber \\
    \textrm{s.t.} &\quad \textrm{tr}(\mathbf{W}^H\mathbf{W}) \leq M
\end{align}
where $\alpha_i$ denotes the rate weight for UE $i$. 

\section{Proposed Deep Neural Network}
\subsection{Deep Neural Network (DNN) Architecture}
\begin{figure}[t]
 \centerline{\includegraphics[width=1\linewidth]{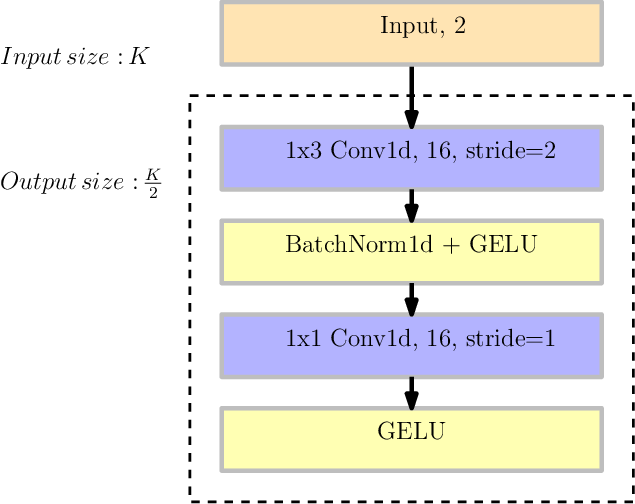}}
  \caption{Basic block structure (dashed part) with 2 input channels and 16 output channels, $\mathrm{BB (2,16)}$.}
  \label{BasicBlock}
  \vspace*{-0.4cm}
\end{figure}

In this section, we present a DNN architecture to design beamforming weights for the sum-rate maximization problem given in (\ref{sum-rate maximizatiob problem}). The input is the frequency domain channel response $\mathbf{H}$ and the output is the beamforming weights $\mathbf{W}$ as introduced in the system model. The backbone of the proposed neural network is composed of convolutional layers followed by batch normalization and activation layers, which are denoted as basic block (BB) together as shown in Fig.~\ref{BasicBlock}.

The convolutional layers operate on the frequency domain information derived through Fourier transform of channel tap data. Flat fading over time slots is assumed since the maximum Doppler shift is set at 10 Hz. Therefore, alterations in channel coefficients are confined to differences across subcarriers. We employ a CNN in our basic block structure, in which 1D convolutions operate on the frequency dimension. To make the input data shape compatible, we reshape it as $\left(BNM,2,K\right)$, where $B$ stands for the batch size of MU-SIMO channel matrices, and the depth dimension represents the IQ samples, while $K$ represents the number of frequency components. 

Batch normalization is used to provide faster convergence and less sensitivity to the initialization of the network parameters. GELU is used as activation function, which has been shown that it provides performance improvement compared to RELU and ELU activations across different tasks such as computer vision, natural language processing, and speech tasks \cite{hendrycksGELU}. Additionally, we increase the number of channels while halving the size of the feature map in basic block structure. Considering the local correlations of physical channels in frequency domain, increase in the depth of the network provides better latent space representation while focusing on more concentrate analysis of local characteristics. This approach is widely adopted in computer vision tasks through popular model architectures to enhance non-linearity facilitating the capture of complex data relationship \cite{he2016deep,simonyan2015very}.  

\begin{figure}[t]
 \centerline{\includegraphics[width=1\linewidth]{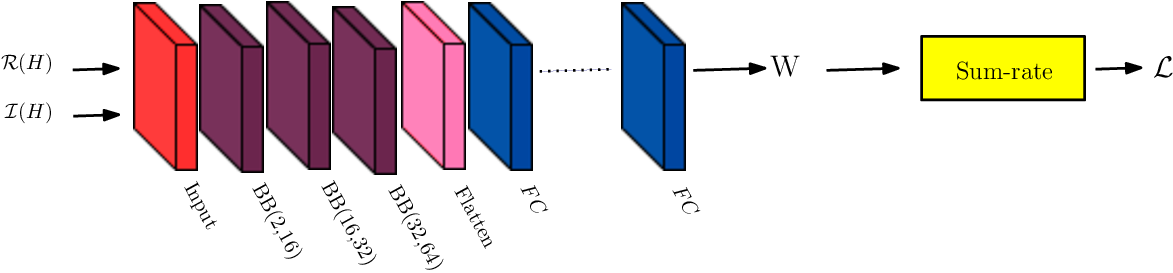}}
  \caption{Deep neural network architecture.}
  \label{DNN_architecture}
  \vspace*{-0.4cm}
\end{figure}

DNN architecture is illustrated in Fig.~\ref{DNN_architecture} by the concatenation of basic block structures, subsequently fully connected (FC) layers. These blocks are characterized by prespecified input and output channel quantities. Flatten layer changes the output shape by concatenating depth dimension for all antenna pairs $(n,m)$, where  $n=1,\ldots,N$ and $m=1,\ldots,M $. Then, the input shape of the first FC layer is $(B,8NMK)$ as shown in Fig.~\ref{DNN_architecture}. The network output after FC layers are reshaped to have beamforming weights $\mathbf{W}$. Furthermore, power normalization is performed for each recevie antenna to satisfy the power constraint given in (\ref{sum-rate maximizatiob problem}).

\subsection{Training Procedure}
The proposed learning procedure offers unsupervised training. The aim is to maximize the sum-rate across all UEs. Therefore, the loss function is specified according to the sum-rate maximization problem given in (\ref{sum-rate maximizatiob problem}),
\begin{align} \label{loss_function}
    \mathcal{L}(\bm{\theta};\mathbf{H}, \widetilde{\mathbf{W}}_{nn}) = -\sum_{i=1}^N \alpha_i \log(1 + \gamma_i)
\end{align}
where $\bm{\theta}$ denotes the network parameters. Note that loss function can be computed by network input $\mathbf{H}$ and network output $f(\bm{\theta}; \mathbf{H}) = \mathbf{W}_{nn}$, where $f(\cdot)$ is the network function. Then, $\widetilde{\mathbf{W}}_{nn}$ represents the power normalized network outputs according to the power constraint. 

For performance evaluation of the proposed network $\mathbf{W}_{nn}$, ZFBF weights $\mathbf{W}_{zf}$ and MMSE beamforming weights $\mathbf{\mathbf{W}}_{mmse}$ are considered as baseline techniques, which can be derived from the channel $\mathbf{H}$ and the noise variance $\sigma^2$ as,
\begin{align}\label{zfbf_formula}
    \mathbf{W}_{zf} &= \left(\mathbf{H}^H\mathbf{H}\right)^{-1}\mathbf{H}^H \\
    \mathbf{W}_{mmse} &= \left(\mathbf{H}^H\mathbf{H}+\sigma^2 \mathbf{I}_N \right)^{-1} \mathbf{H}^H 
\end{align}

\section{Numerical Results and Analysis}
In our experiments, we assess the performance of the proposed framework against ZFBF and MMSE across varying number of receive antennas for single-antenna UEs. The aim of this assessment is to evaluate both spectral efficiency and computational time complexity with the ability to scale to large dimensional massive systems.

\subsection{System and Dataset Specifications}
Channel responses for dataset generation is created according to the TDL-A channel delay profile specified by 3GPP TR 38.901\cite{3gppTR38901}. For system specifications, we use 4 resource blocks with 12 subcarriers each. As stated in DNN architecture before, maximum Doppler shift is set at 10 Hz by assuming flat fading over time slots. Experiments are conducted for SNR range [-15,35] dB. System specifications are shown in Table~\ref{table:system parameters}.

\begin{table} [h!]
\begin{center}
\resizebox{\columnwidth}{!}{%
\begin{tabular}{| c | c |}
\hline
 \textbf{Parameter} & \textbf{Value} \\ \hline
 Channel delay profile & TDL-A  \\  \hline
Number of resource blocks (RBs) & 4 (48 succarriers)  \\ \hline
Delay spread & 30  ns  \\  \hline
Maximum Doppler shift & 10 Hz  \\ \hline
Subcarrier spacing & 30 kHz \\ \hline
Transmission time interval (TTI) & 500 $\mu s$ \\ \hline
SNR & [-15,35] dB   \\\hline
Modulation scheme & QPSK \\ \hline
\end{tabular}}
\end{center}
\caption{System parameters.}
\label{table:system parameters}
\vspace*{-0.5cm}
\end{table}

\subsection{Hyperparameters and Model Specifications}
In our experiments, we choose learning rate as $10^{-4}$. We use a learning rate scheduler which reduces learning rate to half of the current learning rate if the validation loss is not improved for 3 epochs. We use AdamW optimizer. The rate weight  $\alpha_i$ in (\ref{loss_function}) is considered as $\frac{1}{N}$, $ \forall i=1,\ldots, N$. Batch size is taken as 8 when training set and test set have 100 and 25 batches for an epoch, respectively.

\subsection{Results and Analysis}
\begin{figure}[t]
    \begin{center}
     	\subfigure[]{%
     	\includegraphics[width=0.325\linewidth]{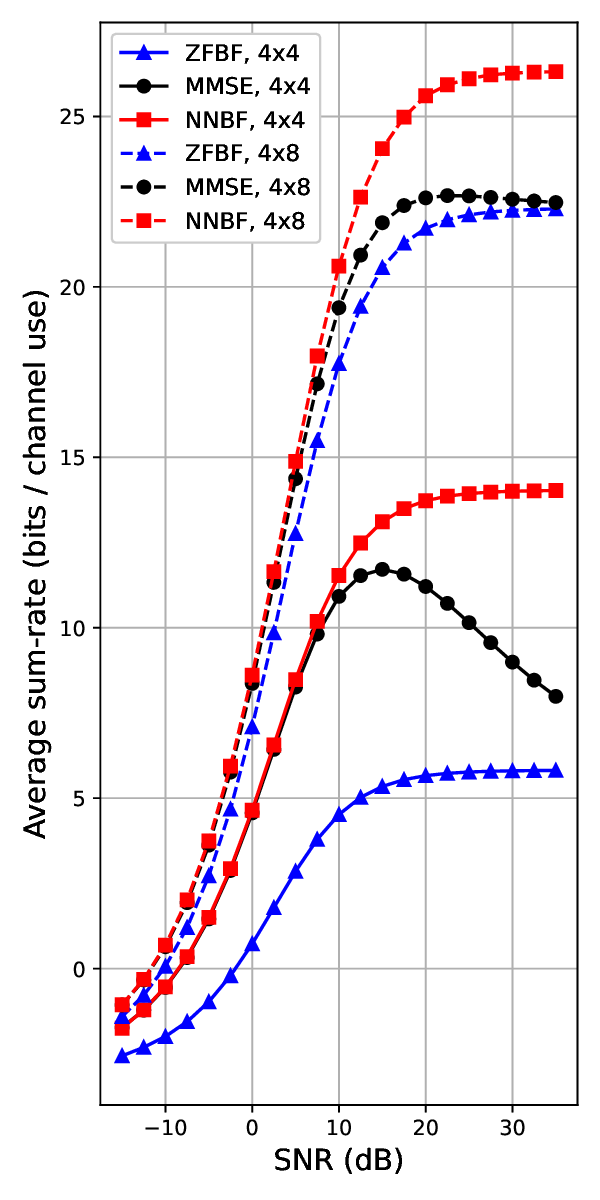}}
     	\subfigure[]{%
     	\includegraphics[width=0.325\linewidth]{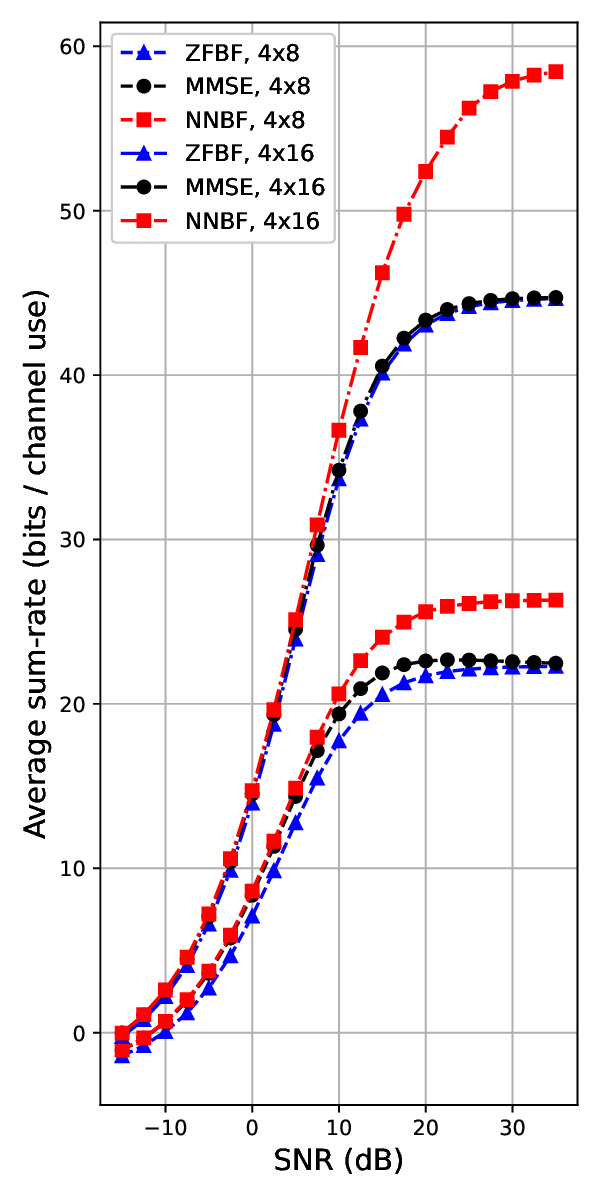}}
     	\subfigure[]{%
     	\includegraphics[width=0.325\linewidth]{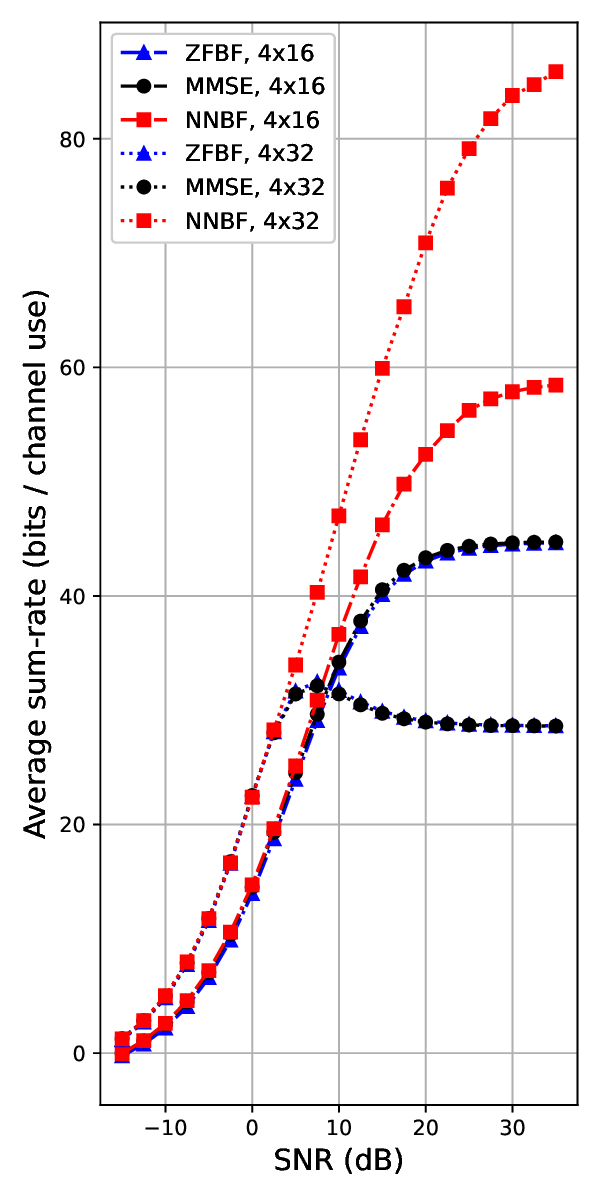}}
    \end{center}
     \caption{Performance comparison of NNBF with baseline methods ZFBF and MMSE, where number of UEs is fixed at 4 and number of receive antennas varies across Rx=\{4,8,16,32\}. Each subplot provides pairwise comparison when number of receive antennas increases: (a) $4\times4$ vs. $4\times8$,  (b) $4\times8$ vs. $4\times16$, (c) $4\times16$ vs. $4\times32$.}
    \label{receive antenna gain impact}
\end{figure}

Fig.~\ref{receive antenna gain impact} illustrates the impact of number of receive antenna for the proposed framework and baseline methods. MMSE, ZFBF, and NNBF results are denoted by black circle, blue triangle, and red square, respectively. For all considered scenarios, the number of single-antenna UEs is fixed at $N=4$. Solid lines, dashed lines, dash-dotted lines and dotted lines represent $M=4,8,16,32$, respectively. In the low SNR regime, NNBF has comparable performance with MMSE while ZFBF has inferior results as can be seen in the comparison of $4\times4$ and $4\times8$ scenarios. As the number of recevie antennas increases, ZFBF can reach the performances of NNBF and MMSE. In high SNR regime, MMSE and ZFBF converge to the same result as expected while the proposed framework NNBF surpasses their performances significantly for all scenarios.
 
\begin{figure}[t]
 \centerline{\includegraphics[width=1\linewidth]{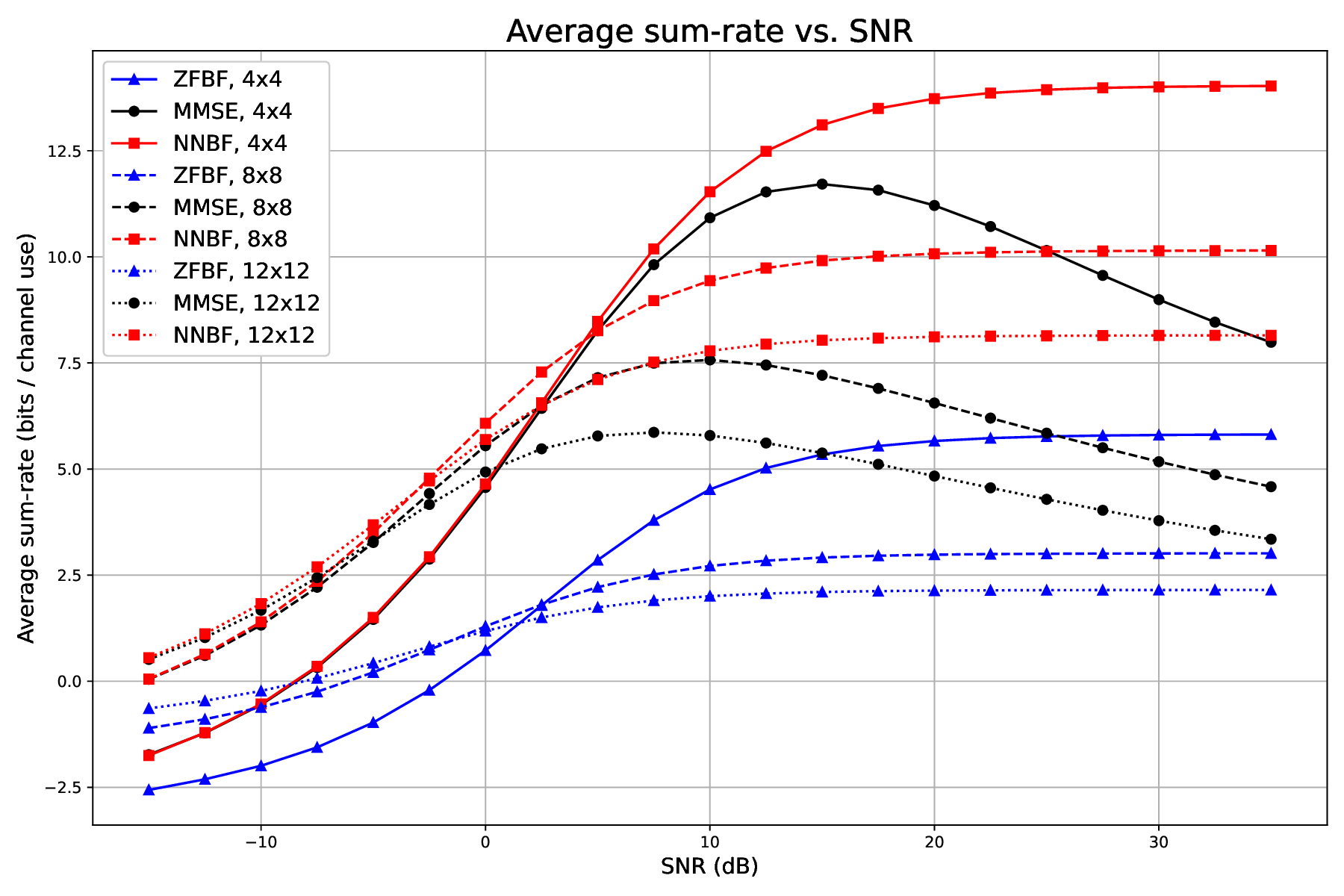}}
  \caption{Average sum-rate vs. SNR for 1:1 ratio of single-antenna UEs to the number of receive antennas.}
  \label{SINR vs. SNR - fixed M/N}
\end{figure}

Fig.~\ref{SINR vs. SNR - fixed M/N} introduces another experimental setup where the proportion of the number of single-antenna UEs to the number of receive antennas remains constant at a 1:1 ratio. Even though a 1:1 ratio may not have any real-world application and 1:4 ratio use cases are much more common, it is worth investigating this configuration for enabling performance comparisons when there are constraints on hardware resources.

In Fig. ~\ref{SINR vs. SNR - fixed M/N}, solid lines represent the scenario for $N=4$ and $M=4$, while dashed lines are for the scenario  $N=8$ and $M=8$ and dotted lines are for $N=12$ and $M=12$. In the low SNR regime, increase in the number of antennas provides better spatial diversity, thereby better mitigation of fading effects. In the high SNR regime, the primary challenge is interference cancellation rather than reducing both noise and interference. Although it is expected that increasing the number of antennas leads to higher spatial degree of freedom to provide better spatial multiplexing gain to cancel interference, more antennas may not provide higher performance necessarily, especially when the ratio is kept fixed. In some scenarios with highly correlated channels, interference cancellation can be more challenging since null steering also degrades the throughput of the desired signal.  Therefore, $4\times4$ configuration has the best performance while $12\times12$ configuration has the worst one in high SNR regime,  which goes against the anticipated trend. Moreover, NNBF performs better for all configurations compared to MMSE and ZFBF across all SNR range. NNBF has also less performance degradation.

\FloatBarrier

\begin{figure}[t]
 \centerline{\includegraphics[width=1\linewidth]{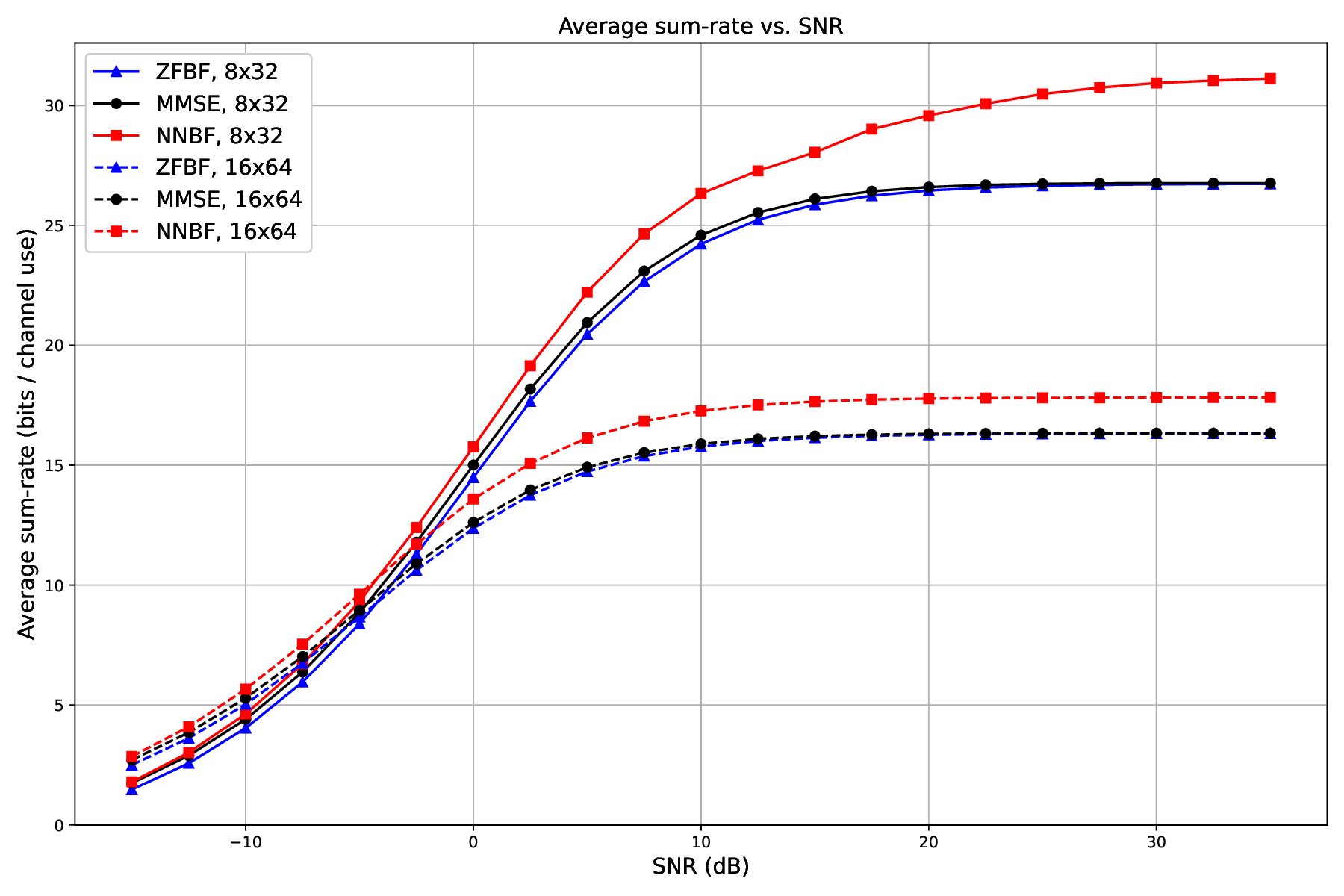}}
  \caption{Average sum-rate versus SNR for a 1:4 ratio of single-antenna UEs to the number of receive antennas.}
  \label{SINR vs. SNR - 32x8 & 64x16}
  \vspace*{-0.4cm}
\end{figure}

Fig.~\ref{SINR vs. SNR - 32x8 & 64x16} illustrates the result of a similar experimental setup where the ratio between the single-antenna UEs and the number of received antennas remains fixed at a 1:4 ratio. This is provided to see whether the results of a 1:1 ratio experiment setup is valid for 1:4 ratio as well. In Fig.~\ref{SINR vs. SNR - 32x8 & 64x16}, solid lines represent the scenario for $8\times32$ configuration while dashed lines represent the scenario for $16\times64$ configuration. Similar to the result of Fig.~\ref{SINR vs. SNR - fixed M/N}, $16\times32$ provides better result in the low SNR regime while $8\times32$ has a better performance in the high SNR regime. NNBF demonstrates better performance than the baseline methods across all SNR range.

\FloatBarrier

\begin{figure}[t]
 \centerline{\includegraphics[width=1\linewidth]{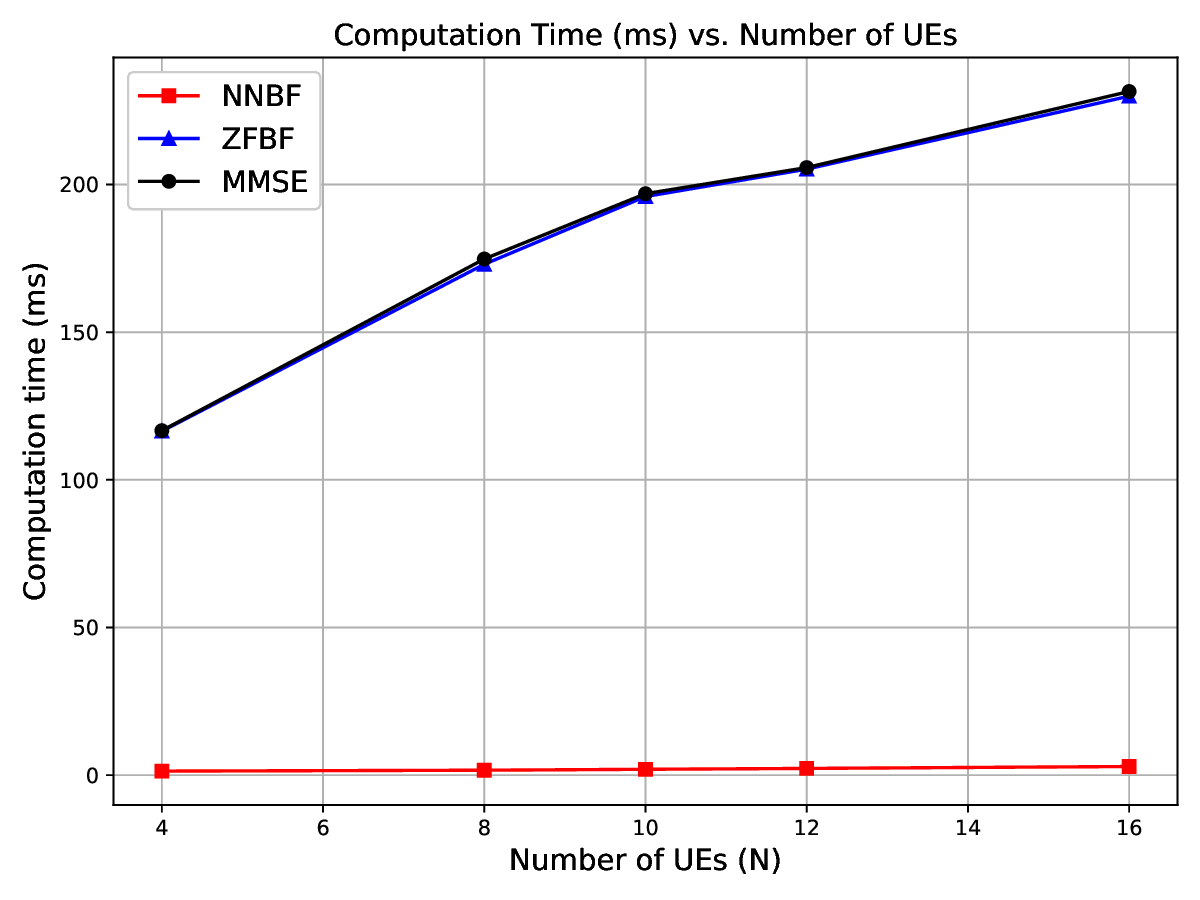}}
  \caption{Computation time (ms) versus number of UEs when the number of receive antennas is 64.}
  \label{computation time}
  \vspace*{-0.4cm}
\end{figure}

Finally, we investigate the computational time of the proposed method compared to the baseline methods MMSE and ZFBF. Fig.~\ref{computation time} shows the computation time across the number of single-antenna UEs when the number of receive antennas is 64. ZFBF and MMSE have similar computation times since pseudo-inverse operation complexity dominates the computation time. NNBF computation time is scalable to the increasing number of UEs with an acceptable increase.

\section{Conclusion}
We proposed a novel unsupervised DL framework that designs uplink receive beamforming for a multi-user SIMO system setting. The proposed frameowkr provides enhanced throughput by maximizing the sum-rate and computational efficiency in contrast to traditional beamforming design techniques of ZFBF and MMSE. Experimental results showed that the proposed NNBF framework outperforms ZFBF and MMSE across a wide range of SNRs including both high and low SNR scenarios. Additionally, NNBF can efficiently scale with an increasing number of single-antenna UEs, unlike ZFBF and MMSE, which involve computationally intensive matrix pseudo-inverse operations.

\bibliographystyle{unsrt}
\bibliography{main}

\begin{thebibliography}{10}

\bibitem{OsheaDL}
T.~O’Shea and J.~Hoydis.
\newblock An introduction to deep learning for the physical layer.
\newblock {\em IEEE Transactions on Cognitive Communications and Networking},
  3(4):563--575, October 2017.

\bibitem{AppML}
Y.~Sun, M.~Peng, Y.~Zhou, Y.~Huang, and S.~Mao.
\newblock Application of machine learning in wireless networks: Key techniques
  and open issues.
\newblock {\em IEEE Communications Surveys \& Tutorials}, 21(4):3072--3108,
  June 2019.

\bibitem{erpek2020deep}
T.~Erpek, T.~J. O'Shea, Y.~E. Sagduyu, Y.~Shi, and T.~C. Clancy.
\newblock Deep learning for wireless communications.
\newblock 2020.
\newblock Available online at arXiv:2005.06068.

\bibitem{ClancyCognitive}
C.~Clancy, J.~Hecker, E.~Stuntebeck, and T.~O'Shea.
\newblock Applications of machine learning to cognitive radio networks.
\newblock {\em IEEE Wireless Communications}, 14(4):47--52, August 2007.

\bibitem{Sun2018}
H.~Sun, X.~Chen, Q.~Shi, M.~Hong, X.~Fu, and N.~D. Sidiropoulos.
\newblock Learning to optimize: Training deep neural networks for interference
  management.
\newblock {\em IEEE Transactions on Signal Processing}, 66(20):5438--5453,
  October 2018.

\bibitem{Hershey2014DeepUnfolding}
J.~R. Hershey, J.~L. Roux, and F.~Weninger.
\newblock Deep unfolding: Model-based inspiration of novel deep architectures.
\newblock 2014.
\newblock Available online at arXiv:21409.2574.

\bibitem{Gregor2010LearningFA}
K.~Gregor and Y.~LeCun.
\newblock Learning fast approximations of sparse coding.
\newblock In {\em ICML}, 2010.

\bibitem{Samuel2017DeepMIMO}
N.~Samuel, T.~Diskin, and A.~Wiesel.
\newblock Deep {MIMO} detection.
\newblock 2017.
\newblock Available online at arXiv:1706.01151.

\bibitem{WenchaoMISODownlinkBF}
W.~Xia, G.~Zheng, Y.~Zhu, J.~Zhang, J.~Wang, and P.~A. Petropulu.
\newblock A deep learning framework for optimization of {MISO} downlink
  beamforming.
\newblock {\em IEEE Transactions on Communications}, 68(3):1866--1880, March
  2020.

\bibitem{BjornsonOptimalResource2013}
E.~Björnson and E.~Jorswieck.
\newblock Optimal resource allocation in coordinated multi-cell systems.
\newblock {\em Foundations and Trends in Communications and Information
  Theory}, 9(2-3):113--381, January 2013.

\bibitem{Huang2018FastBF}
H.~Huang, W.~Xia, J.~Xiong, J.~Yang, G.~Zheng, and X.~Zhu.
\newblock Unsupervised learning-based fast beamforming design for downlink
  {MIMO}.
\newblock {\em IEEE Access}, 7:7599--7605, December 2019.

\bibitem{Zhang_2023}
J.~Zhang, G.~Zheng, Y.~Zhang, I.~Krikidis, and K.~Wong.
\newblock Deep learning based predictive beamforming design.
\newblock {\em IEEE Transactions on Vehicular Technology}, 72(6):8122--8127,
  June 2023.

\bibitem{DeepTx2022}
J.~Huttunen, D.~Korpi, and M.~Honkala.
\newblock {DeepTx}: Deep learning beamforming with channel prediction.
\newblock 2022.
\newblock Available online at arXiv:2202.07998.

\bibitem{Korpi2020DeepRxMC}
D.Korpi, M.~Honkala, J.~Huttunen, and V.~Starck.
\newblock {DeepRx MIMO}: Convolutional {MIMO} detection with learned
  multiplicative transformations.
\newblock In {\em ICC}, 2021.

\bibitem{PowerDL2018}
H.~Ye, G.~Li Ye, and B.~Juang.
\newblock Power of deep learning for channel estimation and signal detection in
  {OFDM} systems.
\newblock {\em IEEE Wireless Communications Letters}, 7(1):114--117, February
  2018.

\bibitem{DeepWaveform2021}
Z.~Zhao, M.~C. Vuran, F.~Guo, and S.~D. Scott.
\newblock Deep-waveform: A learned {OFDM} receiver based on deep complex-valued
  convolutional networks.
\newblock {\em IEEE Journal on Selected Areas in Communications},
  39(8):2407--2420, August 2021.

\bibitem{hendrycksGELU}
D.~Hendrycks and K.~Gimpel.
\newblock Gaussian error linear units ({GELUs}).
\newblock In {\em ICLR}, 2017.

\bibitem{he2016deep}
K.~He, X.~Zhang, S.~Ren, and J.~Sun.
\newblock Deep residual learning for image recognition.
\newblock In {\em CVPR}, 2016.

\bibitem{simonyan2015very}
K.~Simonyan and A.~Zisserman.
\newblock Very deep convolutional networks for large-scale image recognition.
\newblock In {\em ICLR}, 2015.

\bibitem{3gppTR38901}
{3GPP}.
\newblock {Study on channel model for frequencies from 0.5 to 100 GHz}.
\newblock Technical Report TR 38.901, {3rd Generation Partnership Project
  (3GPP)}, April 2022.
\newblock Version 17.0.0.

\end{thebibliography}
\end{document}